# Lead-free, luminescent perovskite nanocrystals obtained through ambient condition synthesis

Fiona Treber[a], Kilian Frank[b], Bert Nickel[b], Carola Lampe*[a], Alexander S. Urban*[a]

[a]  F. Treber, C. Lampe, A. S. Urban
 Nanospectroscopy Group and Center for NanoScience
 Faculty of Physics
 Ludwig-Maximilians-Universität München
 80539 München, Germany
 E-mail: carola.lampe@physik.uni-muenchen.de, Urban@lmu.de

[b]  K. Frank, B. Nickel
 Soft Condensed Matter Group and Center for NanoScience
 Faculty of Physics
 Ludwig-Maximilians-Universität München
 80539 München, Germany

Supporting information for this article is given via a link at the end of the document.

**Abstract:** Heterovalent substitution of toxic lead is an increasingly popular design strategy to obtain environmentally sustainable variants of the exciting material class of halide perovskites. Perovskite nanocrystals (NCs) obtained through solution-based methods exhibit exceedingly high optical quality. Unfortunately, most of these synthesis routes still require reaction under inert gas and at very high temperatures. Herein we present a novel synthesis routine for lead-free double perovskite NCs. We combine hot injection and ligand-assisted reprecipitation (LARP) methods to achieve a low-temperature and ambient atmosphere-based synthesis for manganese-doped $Cs_2NaBiCl_6$ NCs. Mn incorporation is critical for the otherwise non-emissive material, with a 9:1 Bi:Mn precursor ratio maximizing the bright orange photoluminescence (PL) and quantum yield (QY). Higher temperatures slightly increased the material's performance, yet NCs synthesized at room temperature were still emissive, highlighting the versatility of the synthetic approach. Furthermore, the NCs show excellent long-term stability in ambient conditions, facilitating additional investigations and energy-related applications.

## Introduction

Due to their unique optoelectronic properties, halide perovskites have attracted increasing interest over the past decade.[1] Above all, the lead-based representatives of this material class were shown to possess outstanding power conversion efficiencies in thin-film solar cells.[2] At the same time, their nanocrystalline counterparts seem to be ideal for light emission applications.[3] The typically high defect tolerance allows for lead-halide perovskite nanocrystals (NCs) to exhibit photoluminescence quantum yields (PLQY) approaching unity.[3b, 4] At the same time, these NCs are solution processable and offer band gap tunability both via size- and halide composition.[3b, 5] However, the toxicity of lead and the low stability of these compounds in the presence of humidity significantly limit practical applications.[6]

Since the monovalent substitution of lead with its lighter analogs tin or germanium results in even less stable compounds, another strategy gained popularity, employing heterovalent substitution of lead by combining a monovalent and a trivalent cation.[4a, 6b, 7] After the initial report of these double perovskites' potential for optoelectronic applications in 2016, multiple representatives of this material class have been synthesized also as NCs, including the compounds $Cs_2NaBiCl_6$ and $Cs_2AgInCl_6$.[6c, 8] However, their optoelectronic performance still lags far behind the lead-halide perovskites, highlighting the importance of further improvement of the material and research to understand the differences and overcome their shortcomings.[4a, 8e] For that to happen and to facilitate further studies, more simple synthesis methods are necessary, as they provide fast, easy, and cost-efficient access to those materials. Until now, most lead-free double perovskite (LFDP) NCs were synthesized utilizing complex hot injection approaches or similar procedures requiring elevated temperatures and working under an inert atmosphere.[6c, 8b-d, 9]

Here, we report on a facile synthesis method for manganese-doped $Cs_2NaBiCl_6$ (Mn:$Cs_2NaBiCl_6$) NCs under ambient conditions as an alternative to the aforementioned hot injection approach.[8b-d] We find an optimum content of Mn in the precursor material for maximal PLQY and show how repeated purification of the resulting NCs improves optical properties and long-term stability. While the materials' properties are improved with slightly elevated temperatures, the synthesis also performs admirably at room temperature (RT).

## Results and Discussion

The synthesis introduced here combines aspects from both the rapid injection and the anti-solvent precipitation methods often utilized to prepare NCs (ligand-assisted reprecipitation process, LARP).[5a, 10] Yet, it does not require processing under inert gas, and the desired product can be formed at low temperatures, even extending to room temperature. A simplified scheme of the procedure is depicted in Figure 1a. In short, the metal precursors (bismuth(III) acetate, cesium carbonate, manganese(II) acetate, and sodium acetate) were first dissolved in toluene with the help of oleic acid (OA) and oleylamine (OLAm). The mixture was then kept at a constant temperature between RT and 115 °C until trimethylsilyl chloride (TMSCl) was injected, followed by the precipitation with acetone. The colloidal product, redispersed in toluene, exhibits the typical orange-red luminescence under ultraviolet (UV) illumination (Figure 1b), as previously reported.[8b, 8c] The corresponding broad photoluminescence (PL) peak is centered around 595 nm, with the emission dominated by the d-d transition of the Mn$^{2+}$ dopant. On the other hand, the small signal at 400 nm can be assigned to the instrument by comparing it to reference measurements (see Figure S2).

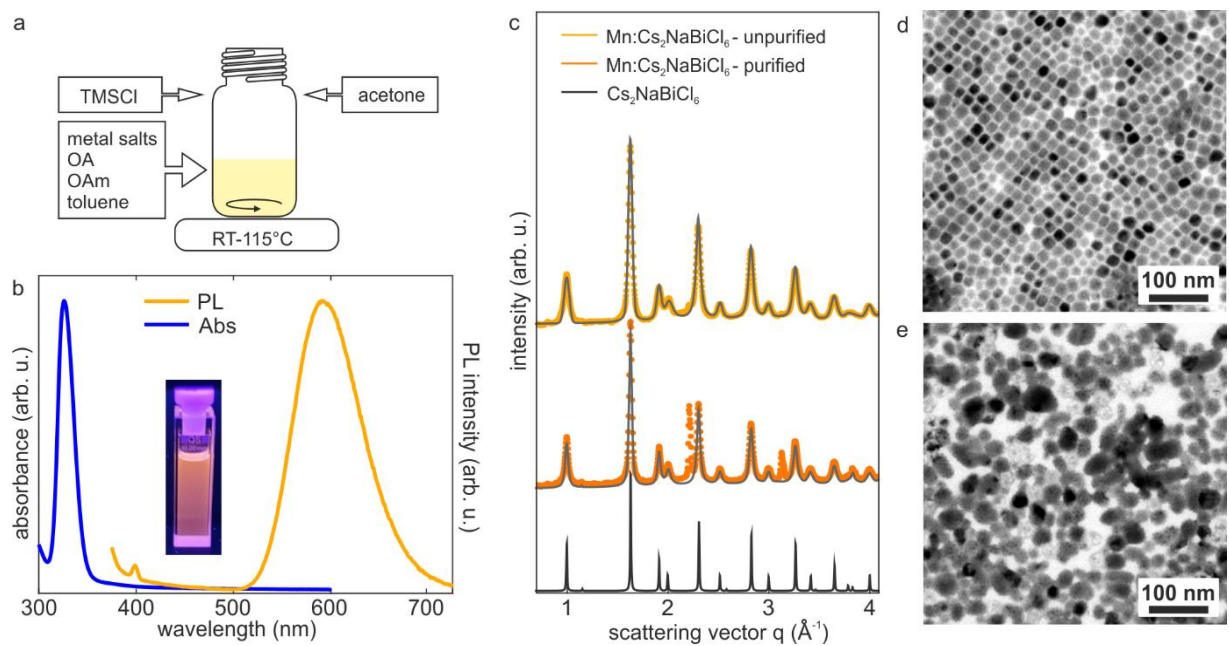

**Figure 1**: (a) Synthesis scheme for an ambient condition procedure for LFDP NCs. (b) Absorption and PL spectra for Mn:Cs$_2$NaBiCl$_6$; inset shows a Mn:Cs$_2$NaBiCl$_6$ sample under UV light illumination. (c) GIWAXS of Mn:Cs$_2$NaBiCl$_6$ (pristine, yellow and purified, orange) and the calculated scattering pattern of Cs$_2$NaBiCl$_6$ (black).[11] TEM images of (d) pristine and (e) purified NCs.

Simultaneously, the absorption spectrum features the characteristic sharp excitonic signal associated with the Bi$^{3+}$ species of the material at around 330 nm.[8b, 8c] To further characterize the synthesized material, wide-angle X-ray scattering experiments were performed under grazing incidence (GIWAXS), and the obtained powder diffractograms are depicted in Fig. 1c. Comparing the experimental pattern with previously reported powder data confirms the formation of the Cs$_2$NaBiCl$_6$ double perovskite with *Fm-3m* space group symmetry.[11] Before and after the purification of the sample, the structure remains virtually unchanged with a lattice constant of 10.8431 Å in excellent agreement with the previously reported value of 10.84290(6) Å. Scherrer analysis revealed a crystallite size of (21.2 ± 4.1) nm before and (30.7 ± 8.4) nm after purification, i.e., the noticeable broadening of the reflexes confirms the nanocrystallinity of the sample, see Supporting Information. In addition to the Cs$_2$NaBiCl$_6$ signal, two reflections (at 2.22 Å$^{-1}$ and 3.14 Å$^{-1}$) occur due to the formation of a degradation product that could not be identified further.

The nanocrystalline nature of the prepared Mn:Cs$_2$NaBiCl$_6$ is further verified by transmission electron microscope (TEM) imaging (Figure 1d). The NCs exhibit a cubic shape with edge lengths ranging from 15 to 20 nm. This morphology is in good accordance with previous publications, where the NC shape was shown to vary with the reaction temperature during the hot injection.[8d] There, temperatures below 170 °C lead to the formation of nanocubes, similar to what we observe here.

The extraction of the NCs from the crude synthesis solution by precipitation and redispersion, as typically done in colloidal semiconductor NC synthesis, was necessary to acquire reliable optical data. The resulting dispersion remains highly absorbent. Repeating this purification procedure up to 3 times does not seem to affect the crystal structure as judged by corresponding GIWAXS patterns (Figure 1c). However, TEM images make it clear that the repeated precipitation and redispersion induce NC aggregation of the NCs (Figures 1d,e). The NCs are visibly larger (20-30 nm) with broader size distribution and irregular shapes. The Scherrer analysis of the X-ray diffraction patterns also confirmed the larger nanocrystal size.

This effect is probably connected to a partial loss of ligands during the purification procedure due to the weak binding of OA and OLAm to the perovskite, which factors in the fusion of individual nanocrystals.[4b] Nevertheless, an increasing number of purification steps has an overall beneficial effect on the optoelectronic properties, as underlined by the trends shown in Figures 2a and S1. The absorbance of the unpurified sample shows a broad feature below 360 nm, while the purified samples exhibit a narrow excitonic absorption feature centered at 330 nm. The PL spectra are very similar, with a broad peak around 595 nm. The PLQY of the pristine sample is below the detection limit of the setup. However, the PLQY increases successively for each washing step, reaching an average maximum of 2.5% after three washing steps (See figure S1). Single measurements did reveal PLQYs up to 4.2%. The addition of small amounts (10 μl) of both OA and OLAm between each purification step (+L), potentially replacing removed ligands, does not seem to result in any significant changes concerning the optoelectronic properties or the morphology.

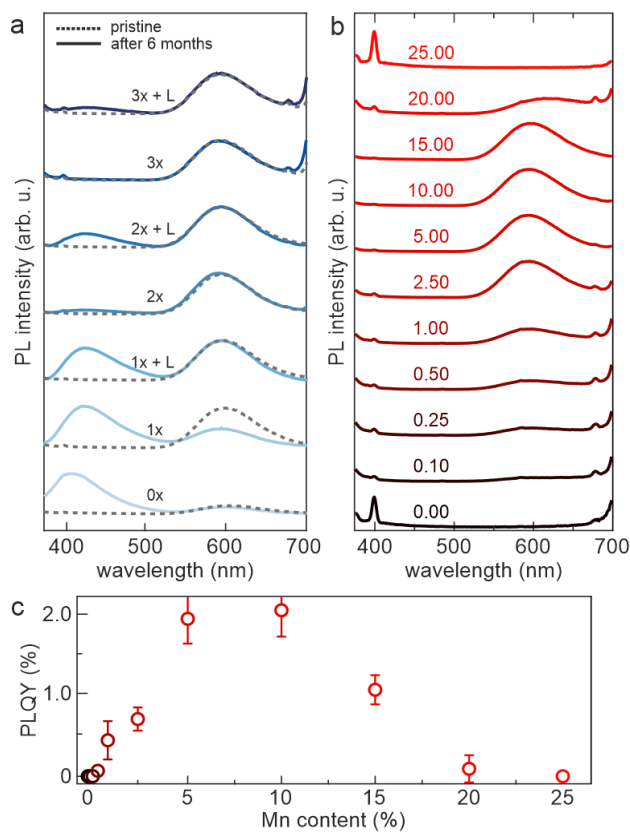

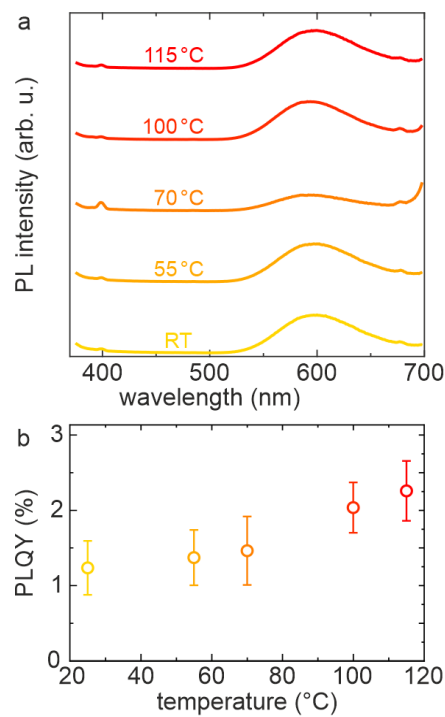

**Figure 2**: (a) PL spectra of Mn:Cs$_2$NaBiCl$_6$ samples after synthesis and after 6 months of storage in ambient conditions with (+L) and without additional ligands during purification steps. (b) PL spectra of Mn:Cs$_2$NaBiCl$_6$ nanocrystals with Mn precursor contents of 0%, 0.1%, 0.25%, 0.5%, 1%, 2.5%, 5%, 10%, 15%, 20%, and 25%. (c) PLQY of Mn:Cs$_2$NaBiCl$_6$ nanocrystals as a function of Mn content.

In contrast, the number of purification steps significantly enhances the long-term stability of the NC dispersions. The three-times washed material exhibits almost identical absorbance and PL signals even after storage for six months under ambient conditions. In their less purified counterparts, another peak emerges in the PL spectrum centered roughly around 400-420 nm. Similar signals were observed in undoped variants of the same composition.[8b, 8c] This signal likely stems from metal oleates in the solution, indicating that the unpurified and weakly purified samples degrade over time (see figure S2).

As the emission of Mn:Cs$_2$NaBiCl$_6$ is based on d-d transitions in manganese as a color center, the amount of Mn in the LFDP is a crucial factor for its optical properties. We show that we can vary the manganese content from 0 to 25% by adequately choosing the metal precursors and analyzing the optical properties, as shown in Figure 2b. Absorption spectra of Mn:Cs$_2$NaBiCl$_6$ exhibit a prominent excitonic peak at around 320 nm for Mn contents up to 20%. For the undoped species, a less pronounced absorption is observed. 25% Mn shows no absorption or PL emission, except an instrument reference signal (discussed before), indicating that the double perovskite NCs do not form at these conditions.

For the samples from 0.10% to 10% Mn, a steady increase of the PL emission around 595 nm is observed. The PL intensity decreases again for even higher Mn contents, indicating a non-ideal stoichiometry in the structure. This behavior can also be quantified by comparing the PLQY of different Mn:Cs$_2$NaBiCl$_6$ samples. The PLQY exhibits a maximum of slightly over 2% for 10% Mn, with a subsequent drop in PLQY for higher Mn concentrations (see figure 2c).

The weak PLQY for small Mn concentrations is expected since the intensity of the observed characteristic Mn d-d transition is directly connected to the Mn amount. Accordingly, light absorbed in the NCs cannot be efficiently transferred to the Mn color centers and is lost via non-radiative decay. The decrease in PLQY for higher Mn concentrations can be ascribed to either the nanocrystals not being able to form correctly - as indicated by the less optimal optical properties of high Mn samples - or inadequate absorption characteristics of Mn:Cs$_2$NaBiCl$_6$. Thus, a manganese concentration of 10% was employed for all subsequent experiments and analyses. GIWAXS measurements of samples at different Mn doping levels were performed to verify the integrity of the double perovskite structure upon Mn doping. The obtained diffractograms are shown in Figure S3. The data show a good agreement with the known cubic *Fm-3m* structure, independent of doping. For the undoped sample, broader diffraction peaks indicate a smaller average crystallite size, (9.7±2.0) nm compared to (25.9±6.8) nm and (27.9±12.4) nm for 15% and 20% doping, respectively. In addition, the effects of crystallite size and microstrain were separated using Williamson-Hall analysis (see Supporting Information). Similar microstrain values of 0.06% to 0.3% are found for the samples prepared with different Mn content, with no clear trend concerning Mn doping.

Ultimately, lead-free double perovskite nanocrystals should be processable at room temperature just like their lead-containing counterparts to enable a widespread application. The synthesis was repeated at different temperatures down to room temperature. Absorption and PL spectra of LFDP NCs synthesized at RT, 55 °C, 70 °C, 100 °C, and 115 °C are comparable for all temperatures, with a slight increase in absorption and emission intensity for increasing temperatures (Figure 3a). This trend can also be seen in the PLQY of the samples, which increases continuously with rising reaction temperature (Figure 3b). While the best results are obtained for moderately elevated temperatures, we demonstrated that this synthesis routine is fully integrable in an ambient atmosphere. An elevated temperature of 140 °C during the precursor dissolution helps shorten the synthesis preparation time, and it is possible to accomplish the entire precursor preparation and the synthesis itself at RT. Additionally, the humidity of the atmosphere was not controlled in the experiments and could therefore vary from synthesis to synthesis. This circumstance further underlines the robustness of the synthetic procedure.

**Figure 3**: PL spectra of for Mn:Cs$_2$NaBiCl$_6$ nanocrystals synthesized with the same composition but different temperatures. (b) Corresponding PLQY values of the Mn:Cs$_2$NaBiCl$_6$ nanocrystal dispersions..

## Conclusion

We have developed a new synthesis routine for fabricating manganese-doped $Cs_2NaBiCl_6$ NCs. The process is inspired by both the hot injection and the LARP processes and was designed to be conducted in ambient atmospheres with moderate temperatures. We found the undoped species $Cs_2NaBiCl_6$ non-emissive, likely due to an indirect bandgap. The introduction of Mn, however, introduces color centers in the structure, and d-d transitions in Mn were observed with a bright orange emission around 595 nm. The effect of repeated purification of the NCs was analyzed and shown to negatively affect the morphology of the NCs while enhancing the PLQY of the samples. The effect of manganese content on the optical properties was tested and found to have an optimum with the highest PLQY at 10% Mn.
Further, we showed that the synthesis could be conducted at room temperature, with the resulting dispersions displaying slightly poorer optical properties. Unpurified, the samples exhibit a slow degradation into different precursor species. However, consecutive purification of the sample after synthesis suppresses the degradation. Further research should be invested in optimizing the emission properties and enhancing the samples' PLQY further. Alternatively, future studies should look into charge separation in these NCs, a necessary pre-requisite, e.g., for photocatalysis.[12]

## Experimental Section

**Materials.** Acetone (99.98%, Merck), bismuth(III) acetate ($BiOAc_3$, 99.999%, ACROS Organics), cesium carbonate ($Cs_2CO_3$, 99%, Sigma-Aldrich), manganese(II) acetate ($MnOAc_2$, 98%, ACROS Organics), oleic acid (OA, 90%, Sigma-Aldrich), oleylamine (OLAm, 70%, Sigma-Aldrich), sodium acetate (NaOAc, 99.997%, Alfa Aesar), toluene (99.8%, VWR Chemicals), trimethylsilyl chloride (TMSCl, MQ100, Merck).

**Mn:$Cs_2NaBiCl_6$ NC synthesis.** Stoichiometric amounts of the metal carbonate and acetate precursors $Cs_2CO_3$ (0.05 mmol, 1eq.), NaOAc (0.05 mmol, 1eq.), $BiOAc_3$ ((0.05−x) mmol, (1−x)eq.) and $MnOAc_2$ (x mmol, x eq.) were dissolved in OA (250 µl) at 140 °C until a transparent solution was obtained. After letting the solution cool down to or below 115 °C, toluene (2 ml) and OLAm (125 µl) were added. The reaction temperature was varied by placing the mixture on a hotplate with the desired temperature for at least 15 min to ensure temperature equalization. Then, the heat source was removed, and TMSCl (0.45 mmol, 9eq.) was injected into the solution under vigorous stirring. After 10 s, acetone (4 ml) was added as the anti-solvent, and the resulting dispersion was stirred for 1 min before centrifuging (10000 rpm, 10 min). The supernatant was discarded. If no further purification was performed, the precipitate was redispersed in toluene (5 ml). Lastly, this dispersion was centrifuged (4000 rpm, 10 min) to remove any potential bulk material, and the supernatant was used for further analysis.

**Purification.** Purification was done via repeated precipitation and redispersion. To this end, the crude residue containing the synthesis product was redispersed in toluene (3 ml) and precipitated again with acetone (6 ml). This procedure comprised one washing step and was repeated up to 3 times before the removal of bulk material/aggregates was carried out, as described in the synthesis section.

**Structural characterization.** For the crystal structure investigation, synchrotron grazing incidence wide-angle X-ray scattering (GIWAXS) measurements were performed at the high energy beamline P07-EH2 (PETRAIII, DESY, Hamburg) with a focused beam.[13] The samples were prepared by drop-casting 80 µl of undiluted colloidal NC solution onto amorphous silica substrates (Microchemicals, 1x1 $cm^2$). The X-ray energy was 103.4 keV (λ = 0.1199 Å) and 103.6 keV (λ = 0.1197 Å) for the data shown in Figure 1 and S4, respectively. A Varex Imaging XRD 4343CT flat panel X-ray detector was used at a 1.301 m (0.7698 m) distance from the sample. Further details are given in the Supporting Information.

**Optical characterization.** The spectroscopic measurements were carried out with a Horiba Fluoromax IV Plus spectrometer equipped with a xenon arc lamp. For the PL and PLQY measurements, the excitation wavelength was set to 355 nm. The PLQY was measured using the integrating sphere Quanta-Phi from Horiba, and several samples were averaged. Additionally, a longpass filter was used to exclude the falsification of the measurements by the remaining excitation light. Therefore, the samples were diluted until their absorbance at the excitation wavelength was below 0.1. All samples were measured in solution (3 ml) using quartz glass cuvettes (Hellma Analytics, optical path length: 1 cm).

## Acknowledgments

The authors gratefully acknowledge support from the Bavarian State Ministry of Science, Research and Arts through the grant "Solar Technologies go Hybrid (SolTech)," from the Deutsche Forschungsgemeinschaft (DFG) under Germany's Excellence Strategy EXC 2089/1-390776260, and the German Ministry for Education and Research (BMBF) through the project "Lucent" (05K19WMA). This work was also supported by the European Research Council Horizon 2020 through the ERC Grant Agreement PINNACLE (759744), the German Ministry for Education, and the Center for NanoScience (CeNS) through a joint research project. We acknowledge DESY (Hamburg, Germany), a member of the Helmholtz Association HGF, for providing experimental facilities. Parts of this research were carried out at PETRA III at DESY, and we would like to thank Ann-Christin Dippel and Olof Gutowski for their assistance in using beamline P07-EH2. Beamtime was allocated for proposal II-20200002.

**Keywords:** lead-free perovskite, synthesis, ambient atmosphere, stability

**Table of Contents Figure**

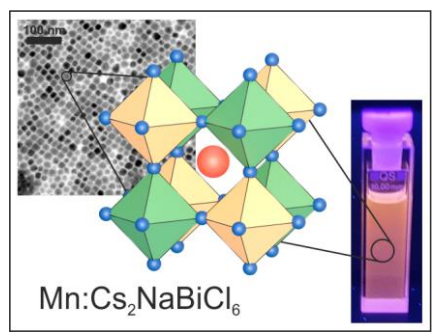